\documentclass[aps,prd,10pt,a4paper,amsfonts,amsmath,amssymb,superscriptaddress,preprintnumbers,showpacs,notitlepage,eqsecnum,nofootinbib]{revtex4-1}

\usepackage{hyperref}
\usepackage{graphicx}

\begin{document}

\title{
Lorentz-violating \textsl{vs} ghost gravitons:
the example of Weyl gravity
}

\author{Nathalie Deruelle}
\affiliation{
APC, CNRS-Universit\'e Paris 7,
75205 Paris CEDEX 13, France
}

\author{Misao Sasaki}
\affiliation{
Yukawa Institute for Theoretical Physics, Kyoto University,
Kyoto 606-8502, Japan
}

\author{Yuuiti Sendouda}
\affiliation{
Graduate School of Science and Technology, Hirosaki University,
Hirosaki, Aomori 036-8561, Japan
}
\affiliation{
APC, CNRS-Universit\'e Paris 7,
75205 Paris CEDEX 13, France
}
\affiliation{
Yukawa Institute for Theoretical Physics, Kyoto University,
Kyoto 606-8502, Japan
}

\author{Ahmed Youssef}
\affiliation{
Institut f\"ur Mathematik und Institut f\"ur Physik,
Humboldt-Universit\"at zu Berlin, 12489 Berlin, Germany
}

\date{
\today
}

\begin{abstract}
We show that the ghost degrees of freedom of Einstein gravity with a Weyl term can be eliminated by a simple mechanism that invokes local Lorentz symmetry breaking.
We demonstrate how the mechanism works in a cosmological setting.
The presence of the Weyl term forces a redefinition of the quantum vacuum state of the tensor perturbations.
As a consequence the amplitude of their spectrum blows up when the Lorentz-violating scale becomes comparable to the Hubble radius.
Such a behaviour is in sharp contrast to what happens in standard Weyl gravity where the gravitational ghosts smoothly damp out the spectrum of primordial gravitational waves.
\end{abstract}

\pacs{04.50.Kd, 04.62.+v, 04.30.-w, 98.80.Cq}

\preprint{YITP-12-3}

\maketitle

\section{
Introduction
}

Four-dimensional gravity theories based on action integrals which are non-linear in the curvature invariants have been part of the landscape of fundamental physics since Weyl \cite{Weyl:1918ib,*Weyl:1919fi,*Weyl:1923} introduced them, soon after Einstein's invention of General Relativity.

Such theories possess, apart from the einsteinian ones, new degrees of freedom (dofs) because the action contains terms which are non-linear in the second derivatives of the metric.
It is commonly believed that, except in $ f(R) $ theories, see e.g.\ \cite{DeFelice:2010aj}, these new dofs will render even flat spacetime unstable, because they are ghosts, that is, dofs with kinetic terms with the ``wrong'' sign, which implies that the phase space of the whole system is no longer finite and its energy is unbounded from below.
The reason why that is considered dangerous is that any kind of interaction term in the action is expected to yield equations of motion whose solutions are erratic, unstable and eventually diverge \cite{Pais:1950za,*Cline:2003gs,*Woodard:2006nt,*Smilga:2004cy,*Smilga:2008pr}.

The simplest representative of such pathological theories is Weyl gravity whose action is Einstein-Hilbert's supplemented by a Weyl-squared term:\footnote{
Units:
$ c = 1 $;
$ \kappa = 8 \pi\,G $\,, $ \gamma $ has dimension $ \text{length}^2 $, $ \kappa $ has dimension $ \text{length}/\text{mass} $, $ \hbar = \kappa\,M_\mathrm{Pl}^2 $\,.
Indices $ a,b,\cdots $ run from $ 0 $ to $ 3 $;
$ i,j,\cdots $ run from $ 1 $ to $ 3 $;
metric $ g_{ab} $ with inverse $ g^{ab} $\,, determinant $ g $ and signature $ (-,+,+,+) $\,.
$ R^a{}_{bcd} = \partial_c \Gamma^a{}_{bd} - \partial_d \Gamma^a{}_{bc} + \cdots $ with $ \Gamma^a{}_{bc} $ the Christoffel symbols;
$ R_{bd} = R^a{}_{bad} $;
$ R = g^{ab}\,R_{ab} $;
$ G_{ab} = R_{ab} - \frac{1}{2}\,g_{ab}\,R $ and $ C_{abcd} = R_{abcd} - \frac{1}{2}\,(g_{ac}\,G_{bd} - g_{ad}\,G_{bc} - g_{bc}\,G_{ad} + g_{bd}\,G_{ac}) - \frac{R}{3}\,(g_{ac}\,g_{bd} - g_{ad}\,g_{bc}) $\,.
}\begin{equation}
S[g_{ab}]
= \frac{1}{2 \kappa}\,\int\!d^4x\,\sqrt{-g}\,
  \left(
   R - \frac{\gamma}{2}\,C_{abcd}\,C^{abcd}
  \right)\,.
\label{eq:actionW}
\end{equation}
It was analysed by Stelle \cite{Stelle:1977ry}, who showed the existence of ghosts (non-tachyonic if $ \gamma > 0 $) but did not exhibit explicitly their malignancy since he worked at linear level, where the dofs do not interact.

In \cite{Deruelle:2010kf}, we analysed linear cosmological perturbations generated during inflation in Weyl gravity, restricting however our attention, as in \cite{Clunan:2009er}, to the case when the Weyl length scale $ \sqrt\gamma $ is shorter than the Hubble scale.
We found that, despite the fact that no interactions between the dofs were taken into account at this linear level, still the scalar modes diverge in the newtonian gauge, although they remain bounded in the comoving slicing.
It therefore seems that the malignancy of ghosts shows up already at linear level, when the background is richer than Minkowski spacetime.

Realising a potential difficulty associated with Weyl gravity already at linear level, we present in the present paper a drastic way out of the potential ``horrors'' of its ghosts.
Following the lines of \cite{Blas:2009yd} and \cite{Germani:2009yt} in their covariant version of Ho\v{r}ava gravity \cite{Horava:2009uw} we introduce a scalar field that breaks local Lorentz covariance (in defining a preferred time direction) and couple it to the Weyl tensor in a way that eliminates its ghosts.
However Lorentz-violation will imply a modification of the dispersion relation of the remaining, einsteinian, dofs.
Specialising to an inflationary background we study tensor perturbations and their spectrum (vector and scalar perturbations are not affected by the presence of our Weyl term) that we compare and contrast to the behaviour of tensor modes and their spectrum in standard Weyl gravity~\eqref{eq:actionW}.

The organisation of the paper is as follows.
In section~\ref{sec:model}, we present our model.
In section~\ref{sec:cosmo} we show, in a cosmological setting, how it eliminates the Weyl ghosts.
Section~\ref{sec:modes} describes the behaviour of tensor perturbations on a de Sitter cosmological background.
In section~\ref{sec:quantum}, we quantise these tensor perturbations to obtain the spectrum of primordial gravitational waves.
Section~\ref{sec:ghost} develops the analysis of \cite{Clunan:2009er} of tensor perturbations in standard Weyl gravity when ghost dofs are present.
Section~\ref{sec:conc} discusses and summarises our results.

\section{\label{sec:model}
The model
}

Our gravity model is specified by the action
\begin{equation}
S[g_{ab},\chi]
= \frac{1}{2 \kappa}\,\int\!d^4x\,\sqrt{-g}\,
  \left(
   R
   + 2 \gamma\,
     C_{abcd}\,C_{efgh}\,\gamma^{ae}\,\gamma^{bf}\,\gamma^{cg}\,u^d\,u^h
  \right)
  + S_\chi[g_{ab},\chi]\,,
\label{eq:action}
\end{equation}
where
\begin{equation}
u_a
\equiv
  \frac{\partial_a \chi}{\sqrt{-\partial_a \chi\,\partial^a \chi}}
\quad\text{and}\quad
\gamma_{ab}
\equiv
  g_{ab} + u_a\,u_b\,.
\end{equation}
Note that the Weyl-term is conformally invariant, see appendix~\ref{app:actions}, where its relation to the Weyl-square scalar is also given.
For our purpose the specific form of $ S_\chi[g_{ab},\chi] $ for the scalar field $ \chi $\, need not be specified, but is requested to be such that it yields solutions in which the gradient vector $ \partial_a \chi $ is everywhere timelike and future-directed.
The vector field $ u^a $ then determines a preferred time direction and that necessarily implies that the theory breaks local Lorentz covariance.
Indeed, the spacetime solution is then preferentially foliated by a family of spacelike hypersurfaces $ \{\Sigma_\chi\} $ on each of which $ \chi $ takes a constant value.
Viewed geometrically, $ u_a $ is the future-directed unit normal to $ \Sigma_\chi $ and $ \gamma_{ab} $ is the induced metric on $ \Sigma_\chi $\,.
The scalar field $ \chi $ can be called a \emph{chronon} as in \cite{Blas:2009yd}.
Note that it can act as an \emph{inflaton} and drive cosmological inflation.

The equations of motion (eoms) for the metric extremise the action with respect to perturbations of the metric and read
\begin{equation}
G_{ab} - \gamma\,B_{ab}
= \kappa\,T^\chi_{ab}\,,
\label{eq:Eeq}
\end{equation}
where $ G_{ab} $ is the Einstein tensor, $ B_{ab} $ is an analogue of the Bach tensor whose expression is given in appendix~\ref{app:eom}, and $ T_\chi^{ab} \equiv 2 (-g)^{-1/2}\,\delta S_\chi/\delta g_{ab} $ is the energy-momentum tensor of $ \chi $\,.
The eom for $ \chi $\,, on the other hand, is
\begin{equation}
\frac{1}{\sqrt{-g}}\,\frac{\delta S_\chi}{\delta\chi}
- \frac{\gamma}{\kappa}\,\nabla_a W^a
= 0\,,
\label{eq:ceq}
\end{equation}
where $ \nabla_a W^a $\,, the variational derivative with respect to $ \chi $ of the Weyl part of the action, is also given in appendix~\ref{app:eom}.

As one can see from the expression of $ B_{ab} $ in appendix~\ref{app:eom}, the Einstein equation~\eqref{eq:Eeq} contains the derivatives of the metric up to fourth order.
One may hence worry that the theory should be pathological because higher order derivatives theories a priori induce ghosts.

However, as a closer examination shows, see its expression in appendix~\ref{app:eom}, $ B_{ab} $ contains only time derivatives up to second order if $ u^a $ is timelike.
Since ghosts are induced by the presence of \emph{time} derivatives higher than the second in the eoms, our theory may therefore be free of ghosts.
Note that our Weyl action does not give rise either to timelike derivatives higher than second in the chronon eom~\eqref{eq:ceq} because the divergence term only contains spacelike derivative of the spacelike vector $ W_a $\,.
The reason why the field equations are not fourth order in time derivatives is due to the presence of the timelike vector $ u^a $\,, which allows to distinguish time and space derivatives, and thus breaks local Lorentz covariance.

\section{\label{sec:cosmo}
Ghost elimination by Lorentz violation
}

Let us first show explicitly and in a specific case that the eoms are indeed second order in time.

We shall consider the example when the constant $ \chi $ surfaces are intrinsically flat and take a flat Friedmann-Lema\^itre-Robertson-Walker (FLRW) spacetime
\begin{equation}
g_{ab}\,dx^a\,dx^b
= a^2(\eta)\,(-d\eta^2 + \delta_{ij}\,dx^i\,dx^j)
\end{equation}
as a background solution, where $ \eta $ is conformal time.

Note that because of conformal flatness of FLRW spacetimes the Weyl-squared term in the action does not affect the background Friedmann equation for the scale factor $ a(\eta) $ since, then, $ B_{ab} $ and $ W_a $ vanish, see appendix~\ref{app:eom};
its evolution is therefore the same as in Einstein gravity coupled to a scalar field $ \chi = \chi(\eta) $\,.

Metric perturbations about a flat FLRW spacetime are decomposed into scalar, vector and tensor parts as
\begin{equation}
\delta g_{ab}\,dx^a\,dx^b
= a^2\,
  \left[
   -2 A\,d\eta^2
   + 2 (\partial_i B + B_i)\,d\eta\,dx^i
   + (2 C\,\delta_{ij}
      + 2 \partial_i \partial_j E
      + \partial_i E_j+ \partial_j E_i
      + h_{ij})\,
     dx^i\,dx^j
  \right]\,,
\label{eq:pert}
\end{equation}
where $ \partial_i B^i = \partial_i E^i = 0 $ and where $ \partial_i h^{ij} = h^i{}_i = 0 $ (all spatial indices being raised with $ \delta^{ij} $).
The following variables are gauge-invariant \cite{Bardeen:1980kt,*Kodama:1985bj,*Mukhanov:1990me}:
\begin{equation}
\Psi
\equiv
  A + (B-E')' + \mathcal H\,(B-E')\,,
\quad
\Phi
\equiv
  C + \mathcal H\,(B-E')\,,
\quad
\Psi_i
\equiv
  B_i -  E'_i\,,
\quad
h_{ij}\,,
\end{equation}
where the Hubble parameter is defined by $ \mathcal H \equiv a'/a $ and where hereafter a prime denotes derivative with respect to $ \eta $\,.

The perturbation to second-order of the action of General Relativity with a scalar field can be expressed in terms of gauge-invariant perturbations of scalar, vector, and tensor types \cite{Bardeen:1980kt,*Kodama:1985bj,*Mukhanov:1990me}.
The perturbation of our higher-curvature term is found to depend on the gauge-invariant tensor and vector variables, $ h_{ij} $ and $ \Psi_i $\,, but not on the scalar variables.
It reads, see \cite{Deruelle:2010kf} and appendix~\ref{app:actions},
\begin{equation}
\begin{aligned}
{}^{(2)}S_{\mathrm W2}[h_{ij},\Psi_i]
&
= \int\!d\eta\,d^3x\,\sqrt{-g}\,
  {}^{(1)}C_{abcd}\,{}^{(1)}C_{efgh}\,
  \gamma^{ae}\,\gamma^{bf}\,\gamma^{cg}\,u^d\,u^h \\
&
= \int\!d\eta\,d^3x\,
  {}^{(1)}C_{ijk0}\,{}^{(1)}C^{ijk}{}_0|_{a(\eta)=1} \\
&
= \frac{1}{2}\,\int\!d\eta\,d^3x\,
  \left(
   \partial_k h'_{ij}\,\partial^k h'^{ij}
   + \frac{1}{2}\,\triangle\Psi_i\,\triangle\Psi^i
  \right)\,,
\end{aligned}
\label{eq:actionW2}
\end{equation}
where $ \triangle \equiv \partial_i \partial^i $\,.

This is where the role of the chronon is unveiled;
it was so coupled to the Weyl tensors that the perturbation of the action only contains first order time derivatives, whereas the other possible combinations, $ C_{abcd}\,C_{efgh}\,\gamma^{ae}\,\gamma^{bf}\,\gamma^{cg}\,\gamma^{dh} $ and $ C_{abcd}\,C_{efgh}\,\gamma^{ae}\,u^b\,u^f\,\gamma^{cg}\,u^d\,u^h $\,, are quadratic in second order time derivatives, see appendix~\ref{app:actions}.

It is clear that the vector perturbations are nondynamical, as no time derivatives of $ \Psi_i $ appear in \eqref{eq:actionW2}.
Moreover they are constrained to vanish just as in pure Einstein gravity.

The total action for the remaining, tensor, perturbations is (see \cite{Grishchuk:1974ny,*Starobinsky:1979ty,*Kodama:1985bj,*Mukhanov:1990me} for the obtention of the Einstein contribution):
\begin{equation}
S_\mathrm T[h_{ij}]
= \frac{1}{8 \kappa}\,\int\!d\eta\,d^3x\,
  \left[
   a^2\,
   (h'_{ij}\,h'^{ij}
    - \partial_k h_{ij}\,\partial^k h^{ij})
   + 4 \gamma\,\partial_k h'_{ij}\,\partial^k h'^{ij}
  \right]\,,
\label{eq:actionh}
\end{equation}
and the corresponding eoms are
\begin{equation}
\left(1 - \frac{4 \gamma\,\triangle}{a^2}\right)\,h''_{ij}
+ 2 \mathcal H\,h'_{ij}
- \triangle h_{ij}
= 0\,.
\label{eq:eomh}
\end{equation}
We hence see explicitly that, as announced, this action contains only Einstein's gravitons as dofs and that the resulting eoms are second order in time derivatives.

Let us now check that these tensorial perturbations are not ghosts.
To do so we decompose them, as usual, into two orthogonal polarisations and go to Fourier space:
\begin{equation}
h_{ij}(\eta,\vec x)
= \sum_{\lambda=1,2} \int\!\frac{d^3k}{(2 \pi)^{3/2}}\,
  e^\lambda_{ij}(\vec k)\,
  h^\lambda_{\vec k}(\eta)\,
  \mathrm e^{\mathrm i\,\vec k \cdot \vec x}\,,
\label{eq:redef}
\end{equation}
where the two polarisation tensors $ e^\lambda_{ij}(\vec k) $\,, such that $ e^\lambda_{ij}(\vec k)\,e_{\lambda'}^{ij}{}^*(\vec k) = \delta^\lambda_{\lambda'} $\,, are transverse and traceless, $ k^j\,e^\lambda_{ij}(\vec k) = e^\lambda{}^i{}_i(\vec k) = 0 $, and where the reality conditions,
\begin{equation}
e^\lambda_{ij}(\vec k)
= e^\lambda_{ij}{}^*(-\vec k)
\quad\text{and}\quad
h^\lambda_{\vec k}(\eta)
= h^\lambda_{-\vec k}{}^*(\eta)\,,
\label{eq:reality}
\end{equation}
are imposed, a star denoting complex conjugation.
Then the action~\eqref{eq:actionh} reads, in Fourier space
\begin{equation}
S_\mathrm T[\{h^\lambda_{\vec k}\}]
= \frac{1}{8 \kappa}\,\sum_{\lambda=1,2} \int\!d\eta\,d^3k\,
  \left[
   (a^2 + 4 \gamma\,k^2)\,|h'^\lambda_{\vec k}|^2
   - a^2\,k^2\,|h^\lambda_{\vec k}|^2
  \right]\,,
\label{eq:actionhF}
\end{equation}
where $ k = |\vec k| $\,.
The graviton can thus be viewed as a collection of non-interacting scalar fields as in General Relativity.
The important point to note here is that $ \gamma $ must be positive, otherwise the graviton modes would be tachyonic ghosts on Minkowski spacetime when $ a(\eta) = 1 $.
Therefore $ (a^2 + 4 \gamma\,k^2) $ is always positive, so that the kinetic terms $ |h'^\lambda_{\vec k}|^2 $ in \eqref{eq:actionhF} are always positive, and the graviton never becomes a ghost on a FLRW background.

\section{\label{sec:modes}
Evolution of tensor perturbations in inflationary cosmology
}

What is to be checked now is whether or not the Weyl term modifies drastically the time evolution of the modes.

The eom in Fourier space deduced from \eqref{eq:eomh} and \eqref{eq:actionhF} is:
\begin{equation}
\chi''_{\vec k} + \Omega_k^2\,\chi_{\vec k}
= 0
\quad\text{with}\quad
\chi_{\vec k}(\eta)
\equiv
  \sqrt{a^2 + 4 \gamma\,k^2}\,h_{\vec k}(\eta)\,,
\end{equation}
where from now on we omit the index $ \lambda $ and where the pulsation $ \Omega_k(\eta) $ is given by\footnote{The analysis of the modes can also be performed in terms of cosmic time $ t $ such that $ dt = a(\eta)\,d\eta $\,, in which case the eom reads $ \ddot f_k + \omega_k^2\,f_k = 0 $ with $ f_k \equiv a^{3/2}\,\sqrt{1 + 4 \gamma\,(k/a)^2}\,h_k $\,, where a dot denotes derivative with respect to cosmic time, where $ H \equiv \dot a/a $ and where
\begin{equation*}
\omega_k^2
= -\frac{1}{4}\,(H^2 + 2 \dot H)
  + \frac{(k/a)^2 - (2 H^2 + \dot H)}{1 + 4 \gamma\,(k/a)^2}
  - \frac{4 \gamma\,(k/a)^2\,H^2}{[1 + 4 \gamma\,(k/a)^2]^2}\,.
\end{equation*}}
\begin{equation}
\Omega_k^2(\eta)
\equiv
  a^2\,
  \left[
   \frac{k^2 - \mathcal H^2 - \mathcal H'}{a^2 + 4 \gamma\,k^2}
   - \frac{4 \gamma\,k^2\,\mathcal H^2}{(a^2 + 4 \gamma\,k^2)^2}
  \right]\,.
\end{equation}

We shall first discuss the time evolution of the modes on a de Sitter background.
Setting
\begin{equation}
a
= \frac{1}{-H\,\eta}
\quad\text{and}\quad
z
\equiv
  -k\,\eta\,,
\end{equation}
$ H $ being a constant, the eom reduces to
\begin{equation}
\frac{d^2\chi_{\vec k}}{dz^2}
+ \frac{z^2 - 2 + 4 \epsilon^2\,z^2\,(z^2 - 3)}
       {z^2\,(1 + 4 \epsilon^2\,z^2)^2}\,
  \chi_{\vec k}
= 0
\quad\text{with}\quad
\chi_{\vec k}
= \frac{k\,\sqrt{1 + 4 \epsilon^2\,z^2}}{H\,z}\,h_{\vec k}\,,
\label{eq:eomchiFdS}
\end{equation}
where we have introduced the dimensionless parameter
\begin{equation}
\epsilon
\equiv
  \sqrt\gamma\,H\,.
\end{equation}
Equation~\eqref{eq:eomchiFdS} can be solved exactly in terms of hypergeometric functions, see next section.
It is  however enlightening to study first the qualitative behaviour of the modes.

As in General Relativity, a mode with wave number $ k $ becomes larger than the horizon at late times, when $ z \ll 1 $, that is
when its physical wavelength $ a/k = 1/(H\,z) $ becomes much larger than the Hubble radius $ H^{-1} $\,.
As for the Lorentz violating regime it holds at early times when $ \epsilon\,z \gg 1 $, that is when the physical wavelength $ a/k $ of the mode $ k $ is still much shorter than the length scale $ \sqrt\gamma $ set by the Weyl term.

Let us first consider the case $ \epsilon \ll 1 $, when the length scale $ \sqrt\gamma $ on which the Lorentz-violating Weyl term operates is much shorter than the Hubble radius $ H^{-1} $\,.
(This is the case if the Weyl correction is viewed as a low energy limit of some Planck scale quantum theory of gravity, inflation happening at, say, the grand unified theory (GUT) scale.)

The modes $ \chi_{\vec k} $\,, which solve \eqref{eq:eomchiFdS}, will then go through three different regimes:
the early, Lorentz violating regime, when $ \epsilon\,z \gg 1 $;
an intermediate regime, $ 1 \gg \epsilon\,z \gg \epsilon $ when they have left the Lorentz violating period but are still sub-horizon;
and the late stage, $ z \ll 1 $ when they have exited the Hubble scale.

In the late and intermediary regimes, when $ \epsilon\,z \ll 1 $, which are Lorentz symmetric, equation~\eqref{eq:eomchiFdS} is the same as in General Relativity:
\begin{equation}
\frac{d^2\chi_{\vec k}}{dz^2}
+ \left(1 - \frac{2}{z^2}\right)\,\chi_{\vec k}
\simeq
  0\,,
\end{equation}
whose two independent solutions are well-known:
after horizon crossing, that is for $ z \ll 1 $, we have $ \chi^{(\mathrm g)}_{\vec k} \propto 1/z $ and $ \chi^{(\mathrm d)}_{\vec k} \propto z^2 $\,, so that the growing tensor perturbation  behaves as $ h_{\vec k} \propto z\,\chi^{(\mathrm g)}_{\vec k} \to \text{const.} $\,, i.e., ``freezes out'';
and on sub-horizon scales $ z \gg 1 $ (but $ z \ll 1/\epsilon $) the modes $ \chi_{\vec k} $ oscillate as $ \sin z $ and $ \cos z $\,.

\begin{figure}
\includegraphics[scale=0.6]{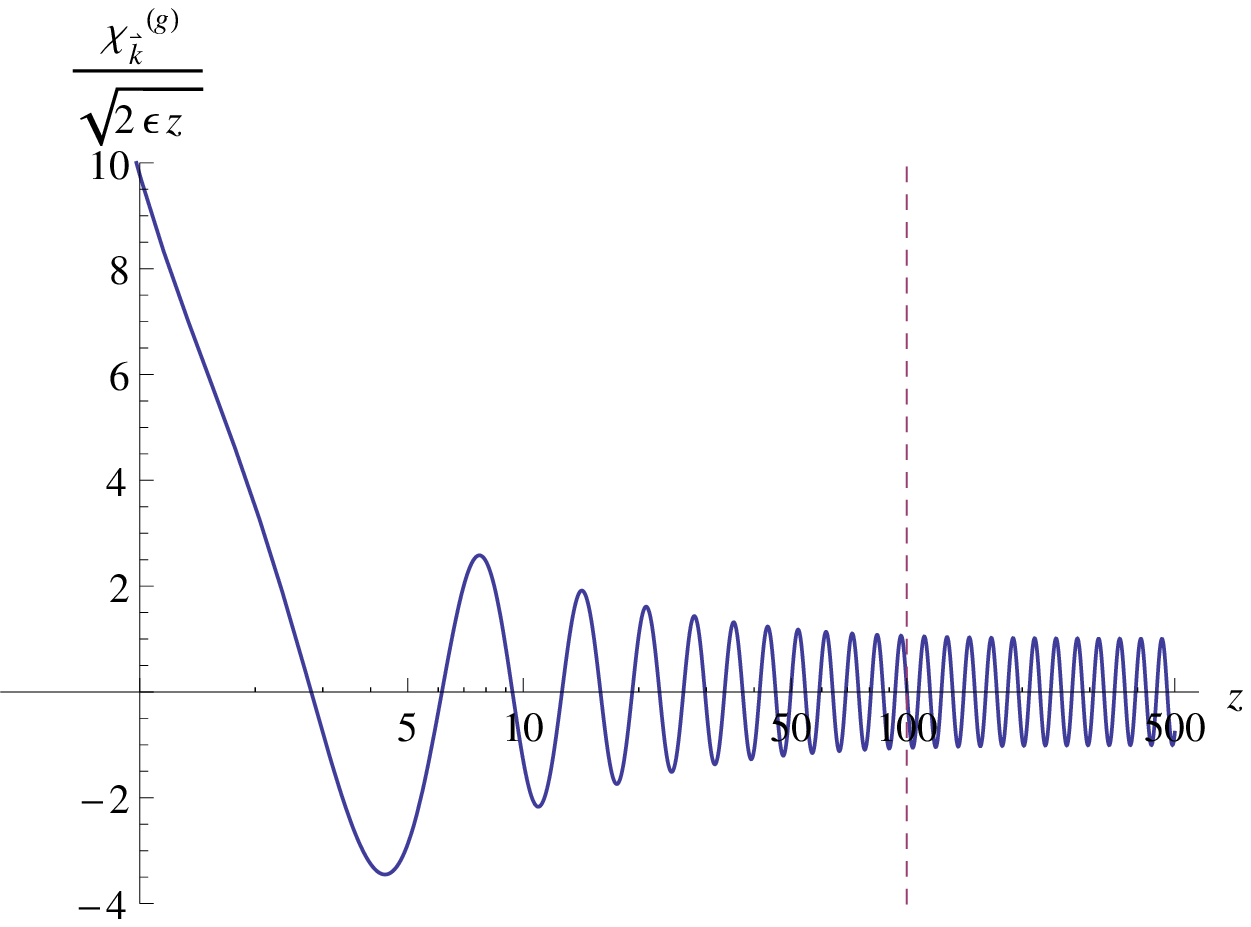}\qquad
\includegraphics[scale=0.6]{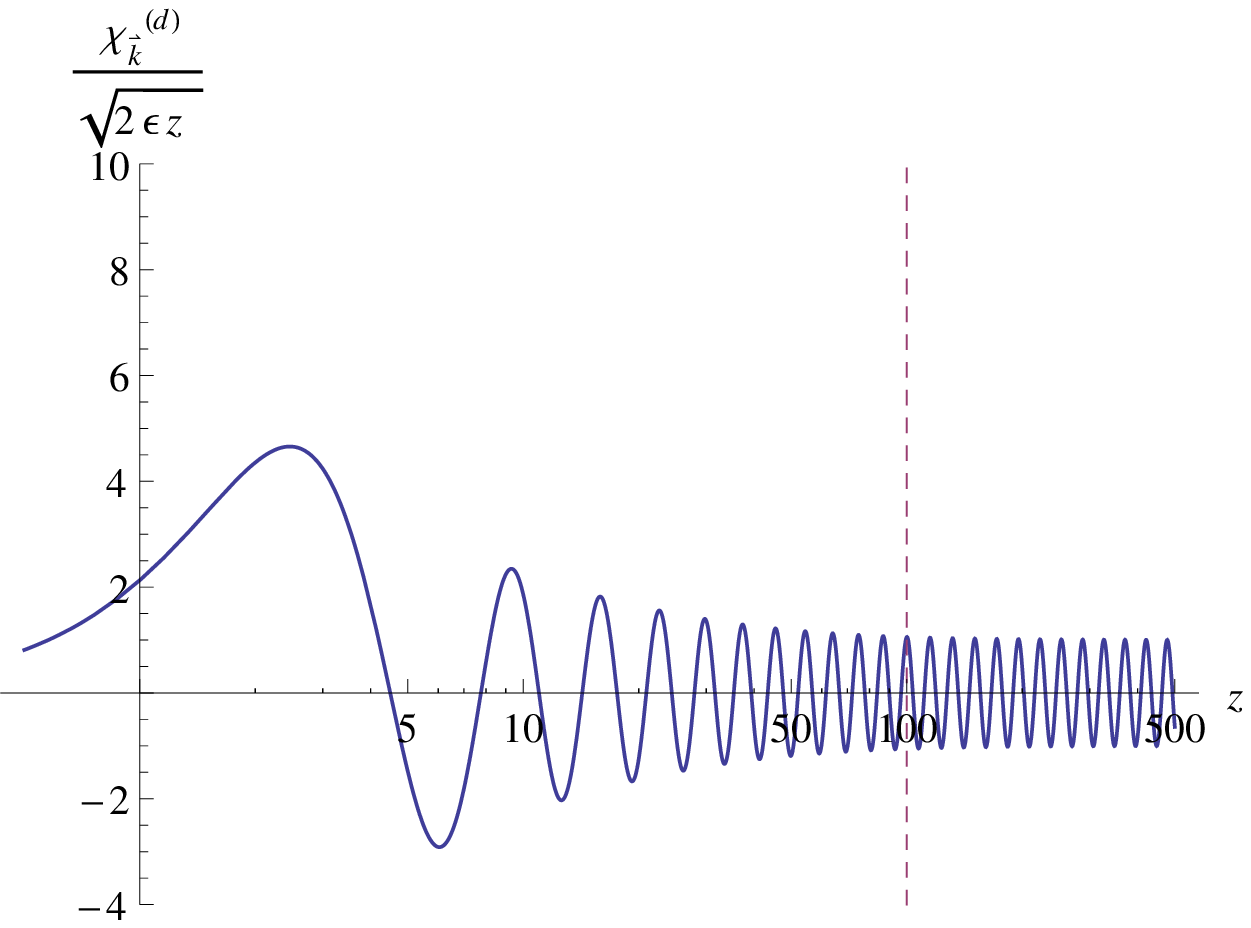}
\caption{\label{fig:chi}
The growing and decaying mode functions $ \chi^{(\mathrm g)}_{\vec k}(z) $ and $ \chi^{(\mathrm d)}_{\vec k}(z) $ on a de Sitter background, with $ z = -k\,\eta = k/(a\,H) $\,, for $ \epsilon \equiv \sqrt\gamma\,H \ll 1 $, that is, when the energy scale $ \kappa\,M_\mathrm{Pl}^2/\sqrt\gamma $ of the Weyl-term is much higher than the inflationary scale $ \kappa\,M_\mathrm{Pl}^2\,H $\,.
In the above figures, where $ \epsilon = 0.01 $ and time increases from right to left, the Lorentz violating period ends at around $ z = 100 $ and Hubble radius crossing occurs at around $ z = 1 $.
}
\end{figure}

In the, early, Lorentz-violating regime, $ \epsilon\,z \gg 1 $\,, $ \chi_{\vec k}\propto h_{\vec k} $\,, and the eom~\eqref{eq:eomchiFdS} reduces to
\begin{equation}
\frac{d^2\chi_{\vec k}}{dz^2}
+ \frac{\chi_{\vec k}}{4 \epsilon^2\,z^2}
\simeq
  0\,,
\end{equation}
whose solutions oscillate as $ \chi_{\vec k} \propto z^{1/2 \pm \mathrm i\,\nu/2} $\,, with $ \nu \equiv \sqrt{1/\epsilon^2 - 1} $ real since $ \epsilon < 1 $.
See figure~\ref{fig:chi}.

Let us now consider the case $ \epsilon \gg 1 $, when the length scale on which the Lorentz-violating Weyl term operates is much \emph{longer} than the Hubble radius $ H^{-1} $\,.
(This would be the case if the Weyl correction was emerging from some effective theory operating below the inflationary, GUT, energy scale.)

At late times, $ \epsilon\,z \ll 1 \ll \epsilon $ the modes behave as before and as in General Relativity: $ \Omega^2_k \simeq -2/z^2 $ so that $ \chi^{(\mathrm g)}_{\vec k} \propto 1/z $ and $ \chi^{(\mathrm d)}_{\vec k} \propto z^2 $\,.
In the intermediary stage on the other hand, when $ 1 \ll \epsilon\,z \ll \epsilon $\,, they are already in the Lorentz-violating stage and since, then, $ \Omega^2_k \simeq -3/(4 \epsilon^2\,z^4) $\,, they behave as $ \chi_{\vec k} \propto z\,\mathrm e^{\pm\sqrt3/(2\,\epsilon\,z)} $ and do not oscillate.
But the really important difference with the previous case occurs at very early times when $ \epsilon\,z \gg \epsilon \gg 1 $.
We still have then that $ \Omega^2_k \simeq 1/(4 \epsilon^2\,z^2) $ but the modes, instead of oscillating, behave as $ \chi_{\vec k} \propto z^{1/2\pm\bar\nu/2} $ with $ \bar\nu = \sqrt{1 - 1/\epsilon^2} $ real.
Because they never oscillate we can qualify these modes as ``rampant'', that is, as ``flourishing and spreading unchecked'' (Oxford dictionary).

We have up to now approximated inflation by a de Sitter stage.
In the case of slow-roll inflation, the Hubble parameter $ H $ is no longer constant but slowly decreases with cosmic time, i.e.\ with decreasing $ z $\,.
The previous results then hold when the parameter $ \epsilon = \sqrt\gamma\,H $ slowly decreases.

Thus, if the energy scale at which inflation starts is lower than the energy scale set by the Weyl term, in other words if the Lorentz-violating length scale is always shorter than the inflationary Hubble radius ($ \epsilon \equiv \sqrt\gamma\,H < 1 $), then all modes behave as in figure~\ref{fig:chi}.
If now inflation starts when the Hubble radius is shorter than the Lorentz-violating scale, then the modes whose wavelengths are big enough to exit the Hubble scale when $ \epsilon $ is still bigger than $ 1 $ will never oscillate.
On the other hand shorter wavelengths modes, which will still be within the Hubble radius when $ \epsilon $ goes through the critical value $ 1 $ will start in a non-oscillatory way at early times and then will behave qualitatively as in figure~\ref{fig:chi}, once $ \epsilon $ becomes smaller than $ 1 $.

Let us summarise sections~\ref{sec:cosmo} and~\ref{sec:modes}:
we confirmed that our Lorentz-violating Weyl action~\eqref{eq:action}, when considered in a cosmological setting, indeed yields second order equations of motion for the tensor perturbations (which are the only ones to be modified by the presence of the Weyl term), see \eqref{eq:eomh};
we confirmed also that, for $ \gamma > 0 $, the perturbations never become ghost-like despite the modification of the coefficient of the kinetic term of the graviton due to the Lorentz-violating Weyl term, see \eqref{eq:actionhF};
finally we studied the perturbation modes, saw that they differ markedly from standard Einstein-de Sitter gravitons in the distant past, and distinguished two different early time behaviours according to whether the Lorentz-violating scale is shorter or longer than the Hubble scale.
These modifications of the early time behaviour of the modes will induce a significant change in the spectrum of quantized primordial gravitational waves, as we shall now see.

\section{\label{sec:quantum}
Quantisation of tensor perturbations and primordial gravitational wave spectrum
}

Let us for clarity write again the action~\eqref{eq:actionh}, for the tensor perturbations on a FLRW background with scale factor $ a(\eta) $\,:
\begin{equation*}
S_\mathrm T[h_{ij}]
= \frac{1}{8 \kappa}\,\int\!d\eta\,d^3x\,
  \left[
   a^2\,(h'_{ij}\,h'^{ij} - \partial_k h_{ij}\,\partial^k h^{ij})
   + 4 \gamma\,\partial_k h'_{ij}\,\partial^k h'^{ij}
  \right]\,.
\end{equation*}
The momentum conjugate to $ h_{ij} $ is
\begin{equation}
\pi^{ij}
= \frac{1}{4 \kappa}\,(a^2\,h'^{ij} - 4 \gamma\,\triangle h'^{ij})\,.
\label{eq:mom}
\end{equation}
Canonical quantisation turns $ h_{ij} $ and $ \pi^{ij} $ into operators satisfying the standard commutation relation
\begin{equation}
[\hat h_{ij}(\eta,\vec x_1),\,\hat{\pi}^{ij}(\eta,\vec x_2)]
= 2 \mathrm i\,\hbar\,\delta(\vec x_1 - \vec x_2)\,,
\label{eq:commcan}
\end{equation}
all other commutators being zero (the factor $ 2 $ comes from the fact that $ \hat h_{ij} $ is the sum of two independent dofs).

Expanding $ \hat h_{ij} $ in Fourier modes as in \eqref{eq:redef}, we further decompose $ \hat h^\lambda_{\vec k} $ as
\begin{equation}
\hat h^\lambda_{\vec k}(\eta)
= \hat a^\lambda_{\vec k}\,h_k(\eta)
  + \hat{\bar a}^\lambda_{-\vec k}\,\bar h_k(\eta)\,,
\end{equation}
where $ \hat a^\lambda_{\vec k} $ and $ \hat{\bar a}^\lambda_{\vec k} $ are arbitrary operators (we introduce $ \hat{\bar a}^\lambda_{-\vec k} $ for later convenience) and where $ h_k(\eta) $ and $ \bar h_k(\eta) $ are two independent solutions, depending on $ k = |\vec k| $ only, of the second order eoms in Fourier space deduced from \eqref{eq:eomh}, that is:
\begin{equation}
\left(1 + \frac{4 \gamma\,k^2}{a^2}\right)\,h''_k
+ 2 \mathcal H\,h'_k
+ k^2\,h_k
= 0\,,
\quad
\left(1 + \frac{4 \gamma\,k^2}{a^2}\right)\,\bar h''_k
+ 2 \mathcal H\,\bar h'_k
+ k^2\,\bar h_k
= 0\,.
\label{eq:eomhF}
\end{equation}
It follows from the eoms~\eqref{eq:eomhF} that the Wronskian of $ h_k(\eta) $ and $ \bar h_k(\eta) $\,,
\begin{equation}
W(h_k,\bar h_k)
\equiv
  h_k\,(a^2 + 4 \gamma\,k^2)\,\bar h'_k
  - \bar h_k\,(a^2 + 4 \gamma\,k^2)\,h'_k\,,
\end{equation}
is a constant $ \neq 0 $ if $ h_k(\eta) $ and $ \bar h_k(\eta) $ are independent solutions of \eqref{eq:eomhF}.
Hence
\begin{equation}
\hat h_{ij}(\eta,\vec x)
= \sum_{\lambda=1,2} \int\!\frac{d^3k}{(2 \pi)^{3/2}}\,
  \left[
   e^\lambda_{ij}(\vec k)\,\hat a^\lambda_{\vec k}\,
   h_k(\eta)\,\mathrm e^{\mathrm i\,\vec k\cdot\vec x}
   + e^\lambda_{ij}{}^*(\vec k)\,\hat{\bar a}^\lambda_{\vec k}\,
     \bar h_k(\eta)\,\mathrm e^{-\mathrm i\,\vec k\cdot\vec x}
  \right]\,,
\label{eq:expansion}
\end{equation}
where we used the reality condition on the polarisation tensors \eqref{eq:reality}.
(Note that we have not yet implemented the hermiticity condition on $ \hat h^\lambda_{\vec k}(\eta) $\,.)

Plugging the expansion~\eqref{eq:expansion} into \eqref{eq:mom}, we find that the commutation relations~\eqref{eq:commcan} are equivalent to
\begin{equation}
[\hat a^\lambda_{\vec k_1},\,\hat{\bar a}^\lambda_{\vec k_2}]
= \delta(\vec k_1-\vec k_2)
\label{eq:commac}
\end{equation}
(all other commutators being zero) if the Wronskian of the two independent mode functions $ h_k $ and $ \bar h_k $ is normalised to
\begin{equation}
W(h_k,\bar h_k)
\equiv
  h_k\,(a^2 + 4 \gamma\,k^2)\,\bar h'_k
  - \bar h_k\,(a^2 + 4 \gamma\,k^2)\,h'_k
= 4 \kappa\,\hbar\,\mathrm i\,.
\label{eq:Wnorm}
\end{equation}

In Minkowski space we have $ a = 1 $ and the eoms~\eqref{eq:eomhF} for the two independent solutions $ h_k(\eta) $ and $ \bar h_k(\eta) $ reduce to
\begin{equation}
\ddot h_k + \omega_{\mathrm Mk}^2\,h_k = 0\,,
\quad
\ddot{\bar h}_k + \omega_{\mathrm Mk}^2\,\bar h_k = 0
\quad\text{with}\quad
\omega_{\mathrm Mk}^2
= \frac{k^2}{1 + 4 \gamma\,k^2}\,,
\end{equation}
where $ \omega^2_{\mathrm Mk} $ is positive and where, for clarity, a dot denotes derivation with respect to cosmic time $ t $ such that $ dt = a(\eta)\,d\eta $\,.

The ``positive frequency'' modes, which will define the vacuum $ |0\rangle $ such that $ \hat a_{\vec k}\,|0\rangle = 0 $ and $ \hat{\bar a}_{\vec k}^\dagger\,|0\rangle = 0 $, are chosen to be
\begin{equation}
h_k
= n_k\,\mathrm e^{-\mathrm i\,\omega_{\mathrm Mk}\,t}\,,
\quad
\bar h_k
= \bar n_k\,\mathrm e^{+\mathrm i\,\omega_{\mathrm Mk}\,t}\,,
\label{eq:modesM}
\end{equation}
where the coefficients $ n_k $ and $ \bar n_k $ must be such that the hermiticity and Wronskian conditions, see \eqref{eq:reality} and \eqref{eq:Wnorm}, are satisfied, that is, such that
\begin{equation}
\hat a^\lambda_{\vec k}\,n_k
= \hat{\bar a}^\lambda_{\vec k}{}^\dagger\,\bar n_k^*\,,
\quad
n_k\,\bar n_k
= \frac{2 \kappa\,\hbar\,\omega_{\mathrm Mk}}{k^2}\,,
\end{equation}
which impose (up to irrelevant constants)
\begin{equation}
\hat a^\lambda_{\vec k}
= \hat{\bar a}^\lambda_{\vec k}{}^\dagger\,,
\quad
n_k
= \bar n_k^*
= \frac{\sqrt{2 \kappa\,\hbar\,\omega_{\mathrm Mk}}}{k}\,.
\end{equation}
Therefore, the choice for the modes is
\begin{equation}
h_k
= \bar h_k^*
= \frac{\sqrt{2 \kappa\,\hbar\,\omega_{\mathrm Mk}}}{k}\,
  \mathrm e^{-\mathrm i\,\omega_{\mathrm Mk}\,t}\,.
\end{equation}
Note for further reference that in the short wavelength limit, $ \gamma\,k^2 \to \infty $\,, the pulsation becomes independent of $ k $\,: $ \omega_{\mathrm Mk} \to (4 \gamma)^{-1/2} $\,, so that we have
\begin{equation}
h_k
\to
  \frac{\sqrt{\kappa\,\hbar}}{k\,\gamma^{1/4}}\,
  \mathrm e^{-\mathrm i\,t/(2 \sqrt\gamma)}\,.
\end{equation}

Next we take accelerated cosmological expansion into account, where another scale $ H $\,, the Hubble parameter, comes into play.

In the de Sitter case when $ a = 1/(-H\,\eta) $ with $ H = \text{const.} $ and $ \eta $ conformal time, the eoms for the modes as given in \eqref{eq:eomhF} read
\begin{equation}
(1 + 4 y^2)\,\frac{d^2\!h_k}{dy^2}
- \frac{2}{y}\,\frac{dh_k}{dy}
+ \frac{h_k}{\epsilon^2}
= 0\,,
\quad
(1 + 4 y^2)\,\frac{d^2\bar h_k}{dy^2}
- \frac{2}{y}\,\frac{d\bar h_k}{dy}
+ \frac{\bar h_k}{\epsilon^2}
= 0\,,
\label{eq:eomhFdS}
\end{equation}
where
\begin{equation}
\epsilon
\equiv
  \sqrt\gamma\,H
\quad\text{and}\quad
y
\equiv
  -\epsilon\,k\,\eta\,.
\end{equation}
As for the Wronskian normalisation condition~\eqref{eq:Wnorm} it becomes
\begin{equation}
h_k\,\frac{d\bar h_k}{dy}
- \bar h_k\,\frac{dh_k}{dy}
= -\mathrm i\,\frac{4 \kappa\,\hbar\,H^2}{k^3\,\epsilon^3}\,
  \frac{y^2}{1 + 4 y^2}\,.
\label{eq:WdS}
\end{equation}

Now, two independent solutions of \eqref{eq:eomhFdS} are
\begin{equation}
h_{(\mathrm g)}(y)
= \frac{1}{2}\,
  F\left(
   \frac{-1-\mathrm i\,\nu}{4},
   \frac{-1+\mathrm i\,\nu}{4},
   -\frac{1}{2};
   -4y^2
  \right)\,,
\quad
h_{(\mathrm d)}(y)
= \frac{32}{3} y^3\,
  F\left(
   \frac{5+\mathrm i\,\nu}{4},
   \frac{5-\mathrm i\,\nu}{4},
   \frac{5}{2};
   -4y^2
  \right)
\label{eq:hghd}
\end{equation}
with
\begin{equation}
\nu
\equiv
\begin{cases}
\sqrt{1/\epsilon^2 - 1}
& \text{if}\quad 0 < \epsilon < 1 \\
\mathrm i\,\sqrt{1-1/\epsilon^2}
& \text{if}\quad \epsilon > 1
\end{cases}\,.
\end{equation}
Note that the hypergeometric functions introduced in \eqref{eq:hghd} are real, whether $ \nu $ is real or imaginary, and that
\begin{equation}
h_{(\mathrm g)}(0)
= \frac{1}{2}\,,
\quad
h_{(\mathrm d)}(0)
= 0\,.
\label{eq:hghd0}
\end{equation}

In order now to determine which combinations of $ h_{(\mathrm g)} $ and $ h_{(\mathrm d)} $ we must choose as our independent modes $ h_k $ and $ \bar h_k $ we have to study the early-time, large $ k $ limit.

\begin{samepage}
When $ y \to \infty $ we have
\begin{equation}
h_{(\mathrm g)}(y)
\to
  c_{(\mathrm g)}^-\,(2 y)^{1/2+\mathrm i\,\nu/2}
  + c_{(\mathrm g)}^+\,(2 y)^{1/2-\mathrm i\,\nu/2}\,,
\quad
h_{(\mathrm d)}(y)
\to
  c_{(\mathrm d)}^-\,(2 y)^{1/2+\mathrm i\,\nu/2}
  + c_{(\mathrm d)}^+\,(2 y)^{1/2-\mathrm i\,\nu/2}
\end{equation}
with, when $ \nu $ is real,
\begin{equation}
c_{(\mathrm g)}^+
= c_{(\mathrm g)}^-{}^*
= -\frac{\sqrt\pi\,\Gamma(-\mathrm i\,\nu/2)}
        {\Gamma^2(-1/4-\mathrm i\,\nu/4)}\,,
\quad
c_{(\mathrm d)}^+
= c_{(\mathrm d)}^-{}^*
= \frac{\sqrt\pi\,\Gamma(-\mathrm i\,\nu/2)}
       {\Gamma^2(5/4-\mathrm i\,\nu/4)}\,.
\label{eq:coeffs}
\end{equation}
(When $ \nu $ is imaginary then the coefficients are real but unequal.)
\end{samepage}

In cosmic time $ t $\,, such that $ dt = a\,d\eta $\,, we have $ a = \mathrm e^{H\,t} $ and $ H\,y = k\,\epsilon\,\mathrm e^{-H\,t} $\,, so that, for $ \nu $ real (i.e.\ for $ 0 < \gamma\,H^2 < 1 $):
\begin{equation}
y^{1/2+\mathrm i\,\nu/2}
\propto
  \mathrm e^{-H\,t/2}\,\mathrm e^{-\mathrm i\,\omega_\mathrm{dS}\,t}
\qquad\text{with}\qquad
\omega_\mathrm{dS}
= \frac{\sqrt{1 - \gamma\,H^2}}{2 \sqrt\gamma}\,,
\end{equation}
which, when $ \gamma\,H^2 \ll 1 $, identifies with the Minkowski positive high frequency modes chosen in \eqref{eq:modesM}\,:
\begin{equation}
y^{1/2+\mathrm i\,\nu/2}
\propto
  \mathrm e^{-\mathrm i\,t/(2 \sqrt\gamma)}\,.
\end{equation}
We recover here what we have discussed in the previous section, to wit that the early time behaviour of the modes differs drastically depending on the value of $\gamma\,H^2 $\,:
they oscillate if $ \nu $ is real ($ \gamma\,H^2 < 1 $), and are ``rampant'' if $ \nu $ is imaginary ($ \gamma\,H^2 > 1 $).
It is clear that quantisation will make (easy) sense only when modes can qualify as ``positive \emph{frequency}'', which implies that they must oscillate.
We shall therefore suppose from now on that
\begin{equation}
0 < \gamma\,H^2 < 1
\quad\Longleftrightarrow\quad
0 < \epsilon \equiv \sqrt\gamma\,H < 1
\quad\Longleftrightarrow\quad
\nu
\equiv
  \sqrt{1/\epsilon^2 - 1}
\quad\text{real}\,.
\end{equation}

In keeping to what we chose when the background is Minkowski spacetime, we shall hence impose that the two independent solutions $ h_k $ and $ \bar h_k $ behave, for large $ y $\,, as
\begin{equation}
h_k(y)
\to
  n_k\,(2 y)^{1/2+\mathrm i\,\nu/2}\,,
\quad
\bar h_k(y)
\to
  \bar n_k\,(2 y)^{1/2-\mathrm i\,\nu/2}\,.
\end{equation}
As for the values of the coefficients $ n_k $ and $ \bar n_k $ they follow from the large $ y $ limit of Wronskian normalisation condition~\eqref{eq:WdS} as well as the hermiticity condition, see \eqref{eq:reality}, which imposes, up to irrelevant constant that
\begin{equation}
\hat a^\lambda_{\vec k}
= \hat{\bar a}^\lambda_{\vec k}{}^\dagger\,,
\quad
n_k
= \bar n_k^*
= \sqrt{\frac{\kappa\,\hbar}{2 \nu\,k^3\,\epsilon^3}}\,H\,.
\end{equation}

Thus, all in all, the hermitian operator $ \hat h_{ij} $ reads
\begin{equation}
\hat h_{ij}(\eta,\vec x)
= \sum_{\lambda=1,2} \int\!\frac{d^3k}{(2 \pi)^{3/2}}\,
  \left[
   e^\lambda_{ij}(\vec k)\,
   \hat a^\lambda_{\vec k}\,
   h_k(\eta)\,
   \mathrm e^{\mathrm i\,\vec k\cdot\vec x}
   + \mathrm{h.c.}
  \right]\,,
\end{equation}
where the modes are:
\begin{equation}
h_k
= \bar h_k^*
= -\mathrm i\,\sqrt{\frac{\kappa\,\hbar\,\nu}{8 k^3\,\epsilon^3}}\,
  H\,(c_{(\mathrm d)}^+\,h_{(\mathrm g)} - c_{(\mathrm g)}^+\,h_{(\mathrm d)})\,,
\label{eq:modes}
\end{equation}
where the functions $ h_{(\mathrm g)} $ and $ h_{(\mathrm d)} $ and the coefficients $ c_{(\mathrm g)} $ and $ c_{(\mathrm d)} $ are defined in \eqref{eq:hghd} and \eqref{eq:coeffs} (and where we used the fact that $ c_{(\mathrm d)}^+\,c_{(\mathrm g)}^- - c_{(\mathrm d)}^-\,c_{(\mathrm g)}^+ = 2 \mathrm i/\nu $).
Finally the ``Bunch-Davies'' vacuum $ |0\rangle $ is defined as usual by
\begin{equation}
\hat a^\lambda_{\vec k}\,|0\rangle
= 0\,.
\end{equation}

The power spectrum $ \mathcal P(k;\eta) $ of the gravitational waves $ h_{ij} $ is now defined as
\begin{equation}
\langle 0|\,\hat h_{ij}(\eta,\vec x_1)\,\hat h^{ij}(\eta,\vec x_2)\,|0\rangle
= \int\!d^3k\,\frac{\mathcal P(k;\eta)}{4 \pi\,k^3}\,
  \mathrm e^{\mathrm i\,\vec k\cdot (\vec x_1 - \vec x_2)}
\end{equation}
and is given by, using the commutation rule~\eqref{eq:commac}\,:
\begin{equation}
\mathcal P(k;\eta)
= \frac{k^3}{\pi^2}\,|h_k(\eta)|^2\,.
\end{equation}

\begin{figure}[ht]
\includegraphics[scale=0.6]{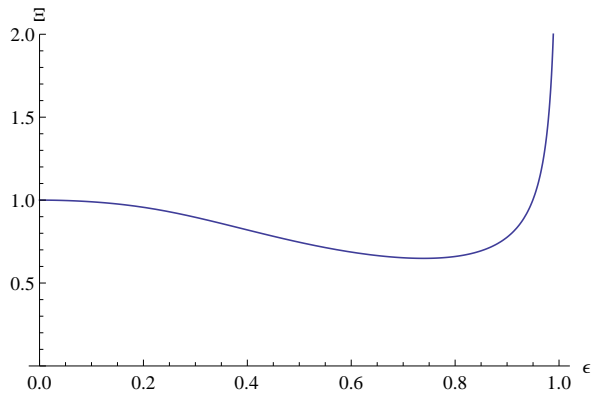}
\caption{\label{fig:Xi}
The modification factor of the power spectrum of the tensor perturbations on a de Sitter background as a function of $ \epsilon = \gamma\,H^2 $\,.
}
\end{figure}

After horizon crossing, that is in the limit $ \eta \to 0 $, we have $ h_k \propto k^{-3/2} $\,, see \eqref{eq:modes}, and the power spectrum is scale invariant just like the power spectrum of ordinary gravitational waves in Einstein theory on a de Sitter background.
However its amplitude depends on $ \nu = \sqrt{1/\epsilon^2-1} $ and is, using the late time limit of $ h_{(\mathrm g)} $ and $ h_{(\mathrm d)} $\,, see \eqref{eq:hghd0}
\begin{equation}
\mathcal P(k;\eta \to 0)
\equiv
  \frac{2\kappa\,H^2}{\pi^2}\,\Xi\,,
\quad\text{where}\quad
\Xi
= \frac{
   \cosh(\pi\,\nu/2)\,
   \coth(\pi\,\nu/2)\,
   \Gamma^2(-1/4+\mathrm i\,\nu/4)\,
   \Gamma^2(-1/4-\mathrm i\,\nu/4)\,
  }{128 \pi^2\,\epsilon^3}\,,
\label{eq:Xi}
\end{equation}
where we used the relations $ \Gamma(-\mathrm i\,\nu/2)\,\Gamma(\mathrm i\,\nu/2) = 2 \pi/[\nu\,\sinh(\pi\,\nu/2)] $ and $ \Gamma(5/4-\mathrm i\,\nu/4)\,\Gamma(5/4+\mathrm i\,\nu/4)\,\Gamma(-1/4-\mathrm i\,\nu/4)\,\Gamma(-1/4+\mathrm i\,\nu/4) = 2 \pi^2/\cosh^2(\pi\,\nu/2) $\,, see figure~\ref{fig:Xi}.

This is the main result of the paper:
In our ghost-free but Lorentz violating Weyl theory of gravity the spectrum of gravitational waves, when evaluated on a de Sitter background, is scale-invariant, as in General Relativity, but its amplitude in the late time limit varies with $ \epsilon \equiv \sqrt\gamma\,H $\,, where $ \sqrt\gamma $ is the length scale on which the Weyl term operates and $ H^{-1} $ is the Hubble radius.
For small $ \epsilon $\,, $ \Xi \approx 1 $:
the spectrum of primordial gravitational waves is almost the same as in General Relativity \cite{Grishchuk:1974ny,*Starobinsky:1979ty} because the Lorentz symmetric regime before horizon crossing lasts long (it spans a $ z $-interval much bigger than $ 1 $);
as $ \epsilon $ increases, that is as the Lorentz-violating length scale approaches the Hubble radius, its amplitude decreases up to a factor $ 65\,\% $, before eventually blowing up as $ \epsilon $ grows further and approaches the critical value $ 1 $.
For $ \epsilon > 1 $ the spectrum, as we saw, cannot be defined as the vacuum expectation value of quantized modes since they do not oscillate at early time.
Another criterion must then be chosen to define it in this regime.

In any case, the spectrum of primordial gravitational waves is greatly modified, if the Lorentz-violating scale happens to be of the same order of magnitude or larger than the Hubble radius.
But, again, it must be stressed that only when the Hubble radius is considerably larger than the Weyl-correction characteristic length scale ($ \epsilon \ll 1 $) do the high frequency modes become those of flat spacetime and it is only then that there is a natural definition of the vacuum state.

In slow-roll inflation now, the Hubble parameter $ H $ is no longer constant but becomes a slowly decreasing function of cosmic time.
The previous results therefore still hold but $ H $ (and hence $ \epsilon = \sqrt\gamma\,H $ and $ \nu = \sqrt{1/\epsilon^2 - 1} $) in the expression~\eqref{eq:Xi} of the spectrum may be evaluated around horizon crossing when $ H(t) = k/a(t) $\,.
This allows to express $ t $\,, and therefore $ H $\,, $ \epsilon $ and $ \nu $ in terms of $ k $\,.
Since $ \epsilon(k) $ is a decreasing function of $ k $\,, the qualitative behaviour of gravitational wave spectrum $ \mathcal P(k;t|_{H=k/a}) $ is then read off from the right to left in figure~\ref{fig:Xi} when $ k $ increases.
What happens when $ \epsilon(k) $ crosses the critical value $ 1 $ during inflation remains to be investigated.

\section{\label{sec:ghost}
Primordial gravitational waves in Einstein plus pure Weyl square gravity
}

In order to compare and contrast the ghost-free but Lorentz-violating Weyl theory of gravity that we have studied in this paper with ordinary ``ghastly''\footnote{i.e.\ ``causing great horror or fear'' (Oxford dictionary)} Weyl theory, we summarise and develop here the results of \cite{Clunan:2009er} to obtain the exact spectrum of gravitational waves on a de Sitter background in Weyl gravity.

When expanded to second order in the tensorial perturbations around a FLRW background, see \eqref{eq:pert} for definitions, the action of Weyl gravity~\eqref{eq:actionW} becomes (see see \cite{Deruelle:2010kf} or appendix~\ref{app:actions})
\begin{equation}
S[h_{ij}]
= \frac{1}{8 \kappa}\,\int\!d\eta\,d^3x\,
  \left[
   a^2\,(h'_{ij}\,h'^{ij} - \partial_k h_{ij}\,\partial^k h^{ij})
   - \gamma\,
     (h''_{ij}\,h''^{ij}
      - 2 \partial_k h'_{ij}\,\partial^k h'^{ij}
      + \partial_{kl} h_{ij}\,\partial^{kl} h^{ij})
  \right]\,,
\end{equation}
which must be compared to \eqref{eq:actionh}.
(Recall, see \cite{Stelle:1977ry}, that $ \gamma $ must be positive to avoid the ghost dofs to be tachyonic on Minkowski spacetime.)
Proceeding as in \cite{Clunan:2009er} ``\`a la'' Ostrogradsky we introduce a new variable $ Q_{ij} \equiv h'_{ij} $ as well as a Lagrange multiplier $ \lambda^{ij} $ (both of them transverse traceless) and consider the equivalent action
\begin{equation}
\begin{aligned}
S[h_{ij},Q_{ij},\lambda^{ij}]
&
= \frac{1}{8 \kappa}\,\int\!d\eta\,d^3x
  \bigl[
   a^2\,(h'_{ij}\,h'^{ij} - \partial_k h_{ij}\,\partial^k h^{ij}) \\
& \quad\quad
   - \gamma\,
     (Q'_{ij}\,Q'^{ij}
      - 2 \partial_k h'_{ij}\,\partial^k h'^{ij}
      + \partial_{kl}\,h_{ij}\,\partial^{kl}\,h^{ij})
   + 2 \lambda^{ij}\,(Q_{ij} - h'_{ij})
  \bigr]\,.
\end{aligned}
\end{equation}
The conjugate momenta are
\begin{equation}
\pi_h^{ij}
= \frac{1}{4 \kappa}\,
  (a^2\,h'^{ij}
   - 2 \gamma\,\triangle h'^{ij}
   - \lambda^{ij})\,,
\quad
\pi_Q^{ij}
= -\frac{\gamma}{4 \kappa}\,Q'^{ij}\,.
\end{equation}
The eoms obtained by extremisation of the action with respect to $ \lambda^{ij} $ and $ Q_{ij} $ are
\begin{equation}
Q_{ij}
= h'_{ij}\,,
\quad
\gamma\,Q''_{ij} + \lambda_{ij}
= 0
\end{equation}
and allow to express the momenta in terms of $ h_{ij}$ alone as
\begin{equation}
\pi_h^{ij}
= \frac{1}{4 \kappa}\,
  (a^2\,h'^{ij} - 2 \gamma\,\triangle h'^{ij} + \gamma\,h'''^{ij})\,,
\quad
\pi_Q^{ij}
= -\frac{\gamma}{4 \kappa}\,h''^{ij}\,.
\label{eq:momW}
\end{equation}
Quantization is implemented by imposing the commutation relations
\begin{equation}
[\hat h_{ij}(\eta,\vec x_1),\,\hat\pi_h^{ij}(\eta,\vec x_2)]
= 2 \mathrm i\,\hbar\,\delta(\vec x_1 - \vec x_2)\,,
\quad
[\hat Q_{ij}(\eta,\vec x_1),\,\hat\pi_Q^{ij}(\eta,\vec x_2)]
= 2 \mathrm i\,\hbar\,\delta(\vec x_1 - \vec x_2)\,,
\label{eq:commW}
\end{equation}
all other commutators being zero and the factor $ 2 $ coming from the fact that both $ h_{ij} $ and $ Q_{ij} $ represent two dofs.

We now expand $ \hat h_{ij} $ in Fourier modes
\begin{equation}
\hat h_{ij}(\eta,\vec x)
= \int\!\frac{d^3k}{(2 \pi)^{3/2}}
  \biggl[
   \biggl(
    \sum_{\lambda=1,2}
    e^\lambda_{ij}(\vec k)\,\hat a^\lambda_{\vec k}\,h^{(1)}_k(\eta)
    + \sum_{\lambda=1,2}
      e^\lambda_{ij}(\vec k)\,\hat b^\lambda_{\vec k}\,h^{(2)}_k(\eta)
   \biggr)\,
   \mathrm e^{\mathrm i\,\vec k\cdot\vec x}
   + \mathrm{h.c.}
  \biggr]\,.
\end{equation}
Plugging this expansion into the definitions of the momenta~\eqref{eq:momW}, we find that the commutation relations~\eqref{eq:commW} are equivalent to imposing (omitting the index $ \lambda $)
\begin{equation}
[\hat a_{\vec k_1},\,\hat a_{\vec k_2}^\dagger]
= \delta(\vec k_1 - \vec k_2)\,,
\quad
[\hat b_{\vec k_1},\,\hat b_{\vec k_2}^\dagger]
= -\delta(\vec k_1 - \vec k_2)
\end{equation}
(all other commutators vanishing) if the mode functions satisfy the following Wronskian conditions
\begin{equation}
\begin{aligned}
&
h^{(1)}_k\,[(a^2 + 2 \gamma\,k^2)\,h'^{(1)}_k{}^* + \gamma\,h'''^{(1)}_k{}^*]
- h^{(2)}_k\,[(a^2 + 2 \gamma\,k^2)\,h'^{(2)}_k{}^* + \gamma\,h'''^{(2)}_k{}^*]
- \mathrm{c.c.}
= 4 \kappa\,\mathrm i\,\hbar\,, \\
&
h'^{(1)}_k\,h''^{(1)}_k{}^* - h'^{(2)}_k\,h''^{(2)}_k{}^*
- \mathrm{c.c.}
= -\frac{4 \kappa\,\mathrm i\,\hbar}{\gamma}\,.
\label{eq:WW}
\end{aligned}
\end{equation}
The constancy of the Wronskians is ensured if the mode functions $ h^{(1)}_k $ and $ h^{(2)}_k $ and their complex conjugates are four independent solutions of the eom for $ h_{ij} $ which reads, in Fourier space
\begin{equation}
(a^2\,h'_k)' + a^2\,k^2\,h_k + \gamma\,(h''''_k + 2 k^2\,h''_k + k^4\,h_k)
= 0\,.
\label{eq:eomW}
\end{equation}

Let us now specialise to a de Sitter background where $ a = 1/(-H\,\eta) $\,, $ H $ being a constant and $ \eta $ conformal time.
Then, as shown in \cite{Clunan:2009er} the eom~\eqref{eq:eomW} factorises neatly:
if $ h^{(1)}_k \equiv z\,\mu^{(1)}_k $ and $ h^{(2)}_k \equiv z\,\mu^{(2)}_k $ are taken to solve, respectively, (with $ z \equiv -k\,\eta $)
\begin{equation}
\begin{aligned}
\frac{d^2\mu^{(1)}_k}{dz^2}
+ \left(1-\frac{2}{z^2}\right)\,\mu^{(1)}_k
= 0\,, \\
\frac{d^2\mu^{(2)}_k}{dz^2}
+ \left(1+\frac{1}{\gamma\,H^2\,z^2}\right)\,\mu^{(2)}_k
= 0\,,
\end{aligned}
\label{eq:eomWbis}
\end{equation}
then, as can be shown explicitly, they satisfy the original eom~\eqref{eq:eomW} and, also, the Wronskian conditions~\eqref{eq:WW} if
\begin{equation}
h^{(1)}_k\,\frac{dh^{(1)}_k{}^*}{dz}
- \mathrm{c.c.}
= h^{(2)}_k\,\frac{dh^{(2)}_k{}^*}{dz}
  - \mathrm{c.c.}
= -\frac{4 \kappa\,\mathrm i\,\hbar\,z^2}{\gamma\,k^3\,[2 + 1/(\gamma\,H^2)]}\,.
\end{equation}
The first equation \eqref{eq:eomWbis} can be seen as the eom for the usual Einstein graviton and the second for the Weyl ghost~dofs.

What remains to be done is to choose ``positive frequency'' modes which define a ``Bunch-Davies'' vacuum state $ |0\rangle $ such that $ \hat a_k\,|0\rangle = \hat b_k\,|0\rangle = 0 $.
This is easy in this case since there is no violation of Lorentz covariance, so that the eoms~\eqref{eq:eomWbis} both reduce to the standard Minkowski eoms for massless fields for $ z \gg 1 $ and $ z \gg 1/(\gamma\,H^2) $\,, that is in the remote past when the physical wavelengths $ a/k $ of the modes are much shorter than both the Hubble radius scale $ H^{-1} $ and the length scale $ \sqrt\gamma $ set by the Weyl term.
We therefore impose that for $ z \to \infty $\,,
\begin{equation}
h^{(1)}_k
\sim
  h^{(2)}_k
\sim
  \sqrt{\frac{2 \kappa}{k^3}}\,
  \frac{H}{\sqrt{1 + 2 \gamma\,H^2}}\,
  z\,\mathrm e^{\mathrm i\,z}\,.
\end{equation}
The solutions of the eoms~\eqref{eq:eomWbis} having this early time/short wavelength behaviour are, up to a phase
\begin{equation}
\begin{aligned}
h^{(1)}_k(z)
&
= \sqrt{\frac{2 \kappa}{k^3}}\,
  \frac{H}{\sqrt{1 + 2 \gamma\,H^2}}\,
  (1 + \mathrm i\,z)\,
  \mathrm e^{\mathrm i\,z}\,, \\
h^{(2)}_k(z)
&
= \sqrt{\frac{2 \kappa}{k^3}}\,
  \frac{H}{\sqrt{1 + 2 \gamma\,H^2}}\,
  \sqrt{\frac{\pi}{2}}\,
  \mathrm e^{\mathrm i \pi\,\bar\nu/2}\,z^{3/2}\,H^{(1)}_{\bar\nu}(z)
\end{aligned}
\end{equation}
with $ \bar\nu \equiv (1/2)\,\sqrt{1-4/(\gamma\,H^2)} $\,.

The power spectrum $ \mathcal P(k;z) $ of the gravitational waves $ h_{ij} $ is again defined as
\begin{equation}
\langle 0|\,\hat h_{ij}(\eta,\vec x_1)\,\hat h^{ij}(\eta,\vec x_2)\,|0\rangle
= \int\!d^3k\,\frac{\mathcal P(k;\eta)}{4 \pi\,k^3}\,
  \mathrm e^{\mathrm i\,\vec k \cdot (\vec x_1-\vec x_2)}
\end{equation}
and is given by
\begin{equation}
\begin{aligned}
\mathcal P(k;z)
&
= \frac{k^3}{\pi^2}\,
  \left(|h^{(1)}_k(z)|^2-|h^{(2)}_k(z)|^2\right) \\
&
= \frac{2 \kappa\,H^2}{\pi^2}\,
  \frac{1}{1 + 2 \gamma\,H^2}\,
  \left(
   1 + z^2
   - \frac{\pi}{2}\,z^3\,
     |\mathrm e^{\mathrm i \pi\,\bar\nu/2}\,H^{(1)}_{\bar\nu}(z)|^2
  \right)\,.
\end{aligned}
\end{equation}
Whatever the value of $ \gamma\,H^2 $\,, that is whatever the sign of $ \bar\nu^2 = \frac{1}{4}\,[1 - 4/(\gamma\,H^2)] $\,, the last two terms do not contribute to the power spectrum when $ z \to 0 $, that is, at late time when the modes have exited the Hubble radius, and we have that
\begin{equation}
\mathcal P(k;z \to 0)
\equiv
  \frac{2 \kappa\,H^2}{\pi^2}\,\Xi_\mathrm W\,,
\quad\text{where}\quad
\Xi_\mathrm W
= \frac{1}{1 + 2 \gamma\,H^2}\,.
\end{equation}
(This generalises \cite{Clunan:2009er}, where the spectrum was computed for small values of the parameter $ \gamma\,H^2 $ only.)

This de Sitter spectrum of primordial gravitational waves at late times, obtained in Einstein plus pure Weyl-square gravity based on the action~\eqref{eq:actionW}, and that obtained in \eqref{eq:Xi} and figure~\ref{fig:Xi} in the ghost-free but Lorentz-violating Weyl gravity theory based on the action~\eqref{eq:action}, are very different: contrarily to $ \Xi $\,, $ \Xi_\mathrm W $ never blows up and, as $ \gamma\,H^2 $ tends to infinity, goes to zero as $ (\gamma\,H^2)^{-1} $\,.

\section{\label{sec:conc}
Conclusion
}

We have proposed and investigated a mechanism to eliminate the ghost degrees of freedom of higher-curvature gravity theories.
The mechanism invokes the gradient of a scalar field which is timelike and therefore implies a breakdown of Lorentz covariance.
Although we only discussed here the particular example of Weyl gravity, the mechanism appears generic enough to eliminate any higher-derivative ghosts.
We expect that a canonical way of analysing generic higher-curvature gravity, developed by the authors \cite{Deruelle:2009zk}, will be useful in studies in this direction.

We investigated quantisation of the inflationary tensor perturbations in our ghost-free Weyl gravity model and defined the ``positive frequency'' modes as those reducing to flat spacetime, Lorentz-violating, positive frequency modes.
Such a modification of the quantum vacuum state may offer an observable signature of Lorentz violation at short wavelengths.
Indeed, we found that in de Sitter inflation this modification of the vacuum state gives rise to an extra overall factor for the super-horizon power spectrum of gravitational-wave background generated from quantum fluctuations.

Since, from a phenomenological point of view, the energy scale of Lorentz violation $ \kappa\,M_\mathrm{Pl}^2/\sqrt\gamma $ can be as low as the Hubble parameter during inflation we expect that inflationary cosmological perturbations may open a useful window to the physics of Lorentz violation.

\begin{acknowledgments}
ND thanks YITP for its enduring and generous hospitality.
She is also grateful to Hirosaki University where this work was completed.
YS thanks ASC at LMU for hospitality, where part of this work was done.
He also thanks Shunichiro Kinoshita for useful conversations.
This work was supported in part by MEXT thorough Grant-in-Aid for Scientific Research (A) No.~21244033 and Grant-in-Aid for Creative Scientific Research No.~19GS0219 (MS) and by JSPS through Postdoctoral Fellowship for Research Abroad and Grant-in-Aid for JSPS Fellows (YS).
This work was also supported in part by MEXT through Grant-in-Aid for the Global COE Program ``The Next Generation of Physics, Spun from Universality and Emergence'' at Kyoto University.
\end{acknowledgments}

\appendix

\section{\label{app:actions}
Weyl-squared actions
}

The Weyl-squared action $ S_{\mathrm W2} = \int\!d^4x\,\sqrt{-g}\,C_{abcd}\,C_{efgh}\,\gamma^{ae}\,\gamma^{bf}\,\gamma^{cg}\,u^d\,u^h $ discussed in the main text is one of the three invariants consisting of quadratic Weyl tensor fully contracted with $ \gamma_{ab} $ and $ u_a $\,;
the other two are
\begin{equation}
\begin{aligned}
S_{\mathrm W0}
= \int\!d^4x\,\sqrt{-g}\,
  C_{abcd}\,C_{efgh}\,\gamma^{ae}\,\gamma^{bf}\,\gamma^{cg}\,\gamma^{dh}\,,
\quad
S_{\mathrm W4}
= \int\!d^4x\,\sqrt{-g}\,
  C_{abcd}\,C_{efgh}\,\gamma^{ae}\,u^b\,u^f\,\gamma^{cg}\,u^d\,u^h\,.
\end{aligned}
\end{equation}
They are such that
\begin{equation}
-4 S_{\mathrm W2} + S_{\mathrm W0} + 4 S_{\mathrm W4}
= \int\!d^4x\,\sqrt{-g}\,C_{abcd}\,C^{abcd}\,,
\end{equation}
hence the coefficient $ -\gamma/2 $ in \eqref{eq:actionW} and $ +2 \gamma $ in \eqref{eq:action}.

A useful fact is that these actions are invariant under a conformal transformation, $ g_{ab} = \Omega^2\,\tilde g_{ab} $\,:
the Weyl tensor is invariant, $ C^a{}_{bcd} = \tilde C^a{}_{bcd} $\,, whereas the geometrical quantities associated with the chronon field are transformed as $ u_a = \Omega\,\tilde u_a $ and $ \gamma_{ab} = \Omega^2\,\tilde \gamma_{ab} $\,, respectively.
Then it can be checked that $ S_{\mathrm W0} $\,, $ S_{\mathrm W2} $ and $ S_{\mathrm W4} $ are all conformally invariant.

The expansion of these actions to second order in gauge-invariant perturbations around FLRW spacetimes are (see main text for definitions and with $ W \equiv \Psi - \Phi $ and \cite{Deruelle:2010kf}):
\begin{equation}
\begin{aligned}
{}^{(2)}S_{\mathrm W2}
&
= -\frac{1}{4}\,\int\!d\eta\,d^3x\,
  (-2 \partial_k h'_{ij}\,\partial^k h'^{ij}
   - \triangle\Psi_i\,\triangle\Psi^i)\,, \\
{}^{(2)}S_{\mathrm W0}
&
= 4\,{}^{(2)}S_{\mathrm W4} \\
&
= \int\!d\eta\,d^3x\,
  \left[
   \frac{1}{4}\,h''_{ij}\,h''^{ij}
   + \frac{1}{4}\,\triangle h_{ij}\,\triangle h^{ij}
   + \frac{1}{2}\,\partial_k h'_{ij}\,\partial^k h'^{ij}
   + \frac{1}{2}\,\partial_i \Psi'_j\,\partial^i \Psi'^j
   + \frac{2}{3}\,(\triangle W)^2
  \right]\,.
\end{aligned}
\end{equation}

\section{\label{app:eom}
Derivation of the equations of motion
}

To compute the variation of our ghost-free Weyl action with respect to the metric, it is useful to employ an ADM-like formalism.
To begin with, we express the action as
\begin{equation}
S_{\mathrm W2}
= \int\!d^Dx\,\sqrt{-g}\,
  C_{abcd}\,C_{efgh}\,\gamma^{ae}\,\gamma^{bf}\,\gamma^{cg}\,u^d\,u^h
\equiv
  \int\!d^Dx\,\sqrt{-g}\,
  W_{abc}\,W^{abc}\,,
\end{equation}
where
\begin{equation}
u_a
\equiv
  N\,\partial_a \chi\,,
\quad
N
\equiv
  \frac{1}{\sqrt{\sigma\,\partial_a \chi\,\partial^a \chi}}\,,
\quad
\gamma_{ab}
\equiv
  g_{ab} - \sigma\,u_a\,u_b\,,
\quad
\sigma
\equiv
  u_a\,u^a\,,
\quad
W_{abc}
\equiv
  \gamma_a{}^d\,\gamma_b{}^e\,\gamma_c{}^f\,u^g\,C_{defg}\,.
\end{equation}
Note that $ W_{[ab]c} = W_{abc} $\,, $ u^a\,W_{abc} = u^a\,W_{bca} = 0 $\,, and $ W_{ab}{}^b = 0 $\,.
We assume $ \sigma $ never vanishes so that the \emph{lapse} function $ N $ remains finite.
We also define the \emph{extrinsic curvature} of the constant-$ \chi $ surfaces by, $\nabla$ being the covariant derivative,
\begin{equation}
K_{ab}
\equiv
  \gamma_a{}^c\,\gamma_b{}^d\,\nabla_d u_c\,.
\end{equation}

The Codazzi relation gives
\begin{equation}
W_{abc}
= \gamma_a{}^d\,\gamma_b{}^e\,\gamma_c{}^f\,
  \left[
   2 \nabla_{[d} K_{e]f}
   - \frac{2}{D-2}\,\nabla_{[d} (K_{e]f} - K\,\gamma_{e]f})
  \right]\,,
\end{equation}
where $ K \equiv \gamma^{ab}\,K_{ab} $\,.
In order to avoid copious appearances of $ \gamma_{ab} $ and $ u_a $\,, we introduce the following notations:
\begin{equation}
\gamma_c{}^d\,\gamma_a{}^e\,\gamma_f{}^b\,\cdots
\nabla_d T_{e\,\cdots}{}^{f\,\cdots}
\to
  D_c T_{a\,\cdots}{}^{b\,\cdots}\,,
\quad
u^d\,\gamma_a{}^e\,\gamma_f{}^b\,\cdots
\nabla_d T_{e\,\cdots}{}^{f\,\cdots}
\to
  \nabla_{\boldsymbol u} T_{a\,\cdots}{}^{b\,\cdots}\,,
\quad
u^a\,T_{a\,\cdots}{}^{b\,\cdots}
\to
  T_{\boldsymbol u\,\cdots}{}^{b\,\cdots}
\end{equation}
and so on.
With this, we can write
\begin{equation}
K_{ab}
= D_a u_b\,,
\quad
W_{abc}
= 2 D_{[a} K_{b]c}
  - \frac{2}{D-2}\,D_{[a} (K_{b]c} - K\,\gamma_{b]c})\,.
\end{equation}
It should be noted that the tensor $ W_{abc} $ only contains one derivative along the direction of $ u_a $\,.

Let us concentrate on the variation with respect to the metric tensor, $ g_{ab} \to g_{ab} + \delta g_{ab} $\,.
Define the variables
\begin{equation}
A
\equiv
  \frac{\sigma}{2}\,u^a\,u^b\,\delta g_{ab}\,,
\quad
B_a
\equiv
  \gamma_a{}^b\,u^c\,\delta g_{bc}\,,
\quad
C_{ab}
\equiv
  \frac{1}{2}\,\gamma_a{}^c\,\gamma_b{}^d\,\delta g_{cd}\,.
\end{equation}
Then we have e.g.\ 
\begin{equation}
\delta g_{ab}
= 2 C_{ab} + 2 \sigma\,u_{(a}\,B_{b)} + 2 \sigma\,A\,u_a\,u_b\,,
\quad
\delta\sqrt{-g}
= C + A\,,
\end{equation}
where $ C \equiv \gamma^{ab}\,C_{ab} $\,.
We also have
\begin{equation}
\delta u_a
= A\,u_a\,,
\quad
\delta u^a
= -A\,u^a - B^a\,,
\quad
\delta\gamma_{ab}
= 2 C_{ab} + 2 \sigma\,u_{(a}\,B_{b)}\,,
\quad
\delta\gamma_a{}^b
= \sigma\,u_a\,B^b\,,
\quad
\delta\gamma^{ab}
= -2 C^{ab}\,,
\end{equation}
and
\begin{equation}
\delta K_{ab}
= \nabla_{\boldsymbol u} C_{ab}
  + 2 K_{(a}{}^c\,C_{b)c}
  - N^{-1}\,D_{(a} (N\,B_{b)})
  - K_{ab}\,A
  + 2 \sigma\,u_{(a}\,K_{b)}{}^c\,B_c\,.
\end{equation}
The variation of the Weyl action with respect to the metric can hence be computed, up to divergences, as
\begin{equation}
\delta S_{\mathrm W2}
= \frac{1}{2}\,\int\!d^Dx\,\sqrt{-g}\,B^{ab}\,\delta g_{ab}\,,
\end{equation}
where
\begin{equation}
\begin{aligned}
B^{ab}
&
\equiv
  4 \nabla_{\boldsymbol u} [N^{-1}\,D_c (N\,W^{c(ab)})]
  + 4 N^{-1}\,D_c (N\,W^{c(ab)})\,K
  - 4 N^{-1}\,D_c (N\,W^{cd(a})\,K_d{}^{b)}
  - 4 N^{-1}\,D_c (N\,W^{c(a|d|})\,K_d{}^{b)} \\
& \quad
  - 4 D_d [N^{-1}\,D_c (N\,W^{cd(a})]\,u^{b)}
  - 4 D_d [N^{-1}\,D_c (N\,W^{c(a|d|})]\,u^{b)}
  + 4 \sigma\,N^{-1}\,D_c (N\,W^{cde})\,K_{de}\,u^a\,u^b \\
& \quad
  + 4 N^{-1}\,D_c (N\,W^{(a|dc|}\,K^{b)}{}_d)
  + 4 N^{-1}\,D_c (N\,W^{cd(a}\,K^{b)}{}_d)
  + 4 N^{-1}\,D_c (N\,W^{d(ab)}\,K^c{}_d) \\
& \quad
  + \frac{8}{D-2}\,W^{c(ab)}\,D^d\,(K_{cd} - K\,\gamma_{cd})
  - 4 W^a{}_{cd}\,W^{bcd}
  - 2 W_{cd}{}^a\,W^{cdb}
  + W^{cde}\,W_{cde}\,g^{ab}\,.
\end{aligned}
\end{equation}
It can be shown that $ B^{ab} $ tensor is traceless and free from the trace part of $ K_{ab} $\,.

A most important property of $ B^{ab} $ is that it contains derivatives $ \nabla_{\boldsymbol u} $ along $ u_a $ up to second order only (in the first term which is of the type $ \nabla_{\boldsymbol u} D W $ with $ W \sim D \nabla_{\boldsymbol u} \gamma $);
hence the eoms remain second order in timelike derivatives whenever $ u_a $ is timelike.
Note also that the  other fourth order terms in $ B_{ab} $ are all of the type $ D D W $ which are first order in time and third order in space when $ u_a $ is timelike.
Because of this distinction between space and time derivatives $ B_{ab} $ is not Lorentz covariant.

The variations of the quantities with respect to $ \chi $ are given by
\begin{equation}
\delta u_a
= V_a\,,
\quad
\delta\gamma_{ab}
= -2 \sigma\,u_{(a}\,V_{b)}
\quad\text{where}\quad
V_a
\equiv
  N\,\gamma_a{}^b\,\partial_b \delta\chi\,.
\end{equation}
The variation of the Weyl action with respect to $ \chi $ is then computed, up to divergence, as
\begin{equation}
\delta S_{\mathrm W2}
= -\int\!d^Dx\,\sqrt{-g}\,
  \nabla_a W^a\,\delta\chi\,,
\end{equation}
where
\begin{equation}
W_a
\equiv
  4 \sigma\,W_{abc}\,C^{b\boldsymbol uc\boldsymbol u}
  + 2 \gamma_{ab}\,W_{cde}\,C^{bcde}\,.
\end{equation}
Note that we may rewrite as $ \nabla_a W^a = N^{-1}\,D_a (N\,W^a) $\,.

\newpage

\bibliographystyle{apsrev4-1}
\bibliography{gravity}

\end{document}